\documentclass[12pt]{article}
 \usepackage{amsmath}
  \usepackage[dvips]{graphicx}

\newcommand{\be}{\begin{equation}}                        %%%%%%%%%
 \newcommand{\ee}{\end{equation}}                         %%%%%%%%%
\newcommand{\bd}{\begin{displaymath}}                     %%%%%%%%%
 \newcommand{\ed}{\end{displaymath}}                      %%%%%%%%%
\newcommand{\bit}{\begin{itemize}}                        %%%%%%%%%
 \newcommand{\eit}{\end{itemize}}                         %%%%%%%%%
\newcommand{\ben}{\begin{enumerate}}                      %%%%%%%%%
 \newcommand{\een}{\end{enumerate}}                       %%%%%%%%%
\newcommand{\baa}{\begin{array}{lll}}                     %%%%%%%%%
 \newcommand{\eaa}{\end{array}}                           %%%%%%%%%
\newcommand{\ba}{\begin{eqnarray}}                        %%%%%%%%%
 \newcommand{\ea}{\end{eqnarray}}                         %%%%%%%%%
 \newcommand{\ro}{\rho}                                   %%%%%%%%%
  \newcommand{\si}{\sigma}                                %%%%%%%%%
\newcommand{\Tr}{\mbox{Tr\,}}                             %%%%%%%%%
\newcommand{\Ds}{\displaystyle}                           %%%%%%%%%
\newcommand{\q}{\bar{q}}                                  %%%%%%%%%
  \newcommand{\qq}[1]{\langle\bar{q}{#1}q\rangle}          %%%%%%%%%
\newcommand{\va}[1]{\langle{#1}\rangle}                   %%%%%%%%%
 \newcommand{\nn}{\nonumber}                              %%%%%%%%%
\newcommand{\gev}[1]{\relax\ifmmode{\text{GeV}^{#1}}      %%%%%%%%%
                     \else{GeV$^{#1}${ }}\fi}             %%%%%%%%%
\newcommand{\Mev}{\relax\ifmmode{\text{MeV}}              %%%%%%%%%
                     \else{MeV{ }}\fi}                    %%%%%%%%% 
\newcommand{\xxyy}                                        %%%%%%%%%
 {\left(\bar x\rightarrow x,\,\bar y\rightarrow y\right)} %%%%%%%%%
\def\MSbar{\relax\ifmmode\overline                        %%%%%%%%%
            {\rm MS}\else{$\overline{\rm MS}${ }}\fi}     %%%%%%%%%
\def\as{\relax\ifmmode \alpha_s\else{$ \alpha_s${ }}\fi}  %%%%%%%%%
\def\abar{\relax\ifmmode{\bar{a}}\else{$\bar{a}${ }}\fi}  %%%%%%%%%
  \def\ie{\hbox{\it i.e.}{ }} %%%%%%%%%
   \def\eg{\hbox{\it e.g.}{ }}  %%%%%%%%%
\newcounter{myfig}                                        %%%%%%%%%
\newcommand{\myfig}{\refstepcounter{myfig}}               %%%%%%%%%
\begin{document}
%\phantom{.}\hfill \today 

\begin{center}
{\Large Lattice measurements of nonlocal quark condensates,\\
         vacuum correlation length, and pion
           distribution amplitude in QCD }\\[0.5cm]
A.~P.~Bakulev\footnote{E-mail: bakulev@thsun1.jinr.ru},\ \
S.~V.~Mikhailov\footnote{E-mail: mikhs@thsun1.jinr.ru}\\[0.5cm]

\textit{Joint Institute for Nuclear Research,
Bogoliubov  Lab. of Theoretical Physics,\\
141980, Moscow Region, Dubna, Russia}\\[0.5cm]

\end{center}
\begin{abstract}
    Recent data of lattice measurements 
    of the gauge-invariant nonlocal scalar quark condensates 
    are analyzed to extract the short--distance correlation length,
    $1/\lambda_q $, 
    and to construct an admissible Ansatz for the condensate behaviour 
    in a coordinate space. 
    The correlation length values for both the quenched 
    and full-QCD cases appear in a good agreement 
    with the well-known QCD SR estimates of the mixed quark-gluon condensate,
    $2\lambda_q^2 = \qq{\left( ig\si_{\mu\nu}G_{\mu\nu}\right)}/\qq{} 
    = 0.8-1.1~\gev{2}$. 
    We test two different Ansatzes for a nonlocal quark condensate
    and trace their influence 
    on the twist-2 pion distribution amplitude 
    by means of QCD SRs. 
    The main features of the pion distribution amplitude 
    are confirmed by the CLEO experimental results.
\end{abstract}
\vspace {10mm}

\vspace {8mm}
%%%%%%%%%%%%%%%%%%%%%%%%%%%%%%%%%%%%%%%%%%%%%%%%%%%%%%%%%%%%%%%%%%%
\newpage

\section{Introduction}
 \label{intro}
%%%%%%%%%%%%%%%%%%%%%%%%%%%%%%%%%%%%%%%%%%%%%%%%%%%%%%%%%%%%%%%%%%%
The results of long--awaited lattice measurements 
of gauge-invariant nonlocal quark condensates 
have been published recently in~\cite{DDM99}. 
These data provide a possibility to examine directly 
the models of nonlocal--condensate coordinate behaviour.
Different models have been suggested in the framework of QCD sum rules (SR)
for dynamic properties of light mesons~\cite{MR86,BR91,MR92,Rad94,BM94}.
We used an extended QCD SR approach 
with nonlocal condensates~\cite{MR92,BR91,BM95,BM98}
as a bridge to connect meson properties,
form factors, and distribution amplitudes 
with the structure of QCD vacuum. 
For light meson phenomenology, the value of short--distance 
correlation length $l$ in QCD vacuum, 
$\Ds l \simeq \frac1{\lambda_q}$,
has a paramount significance,
while the details of a particular condensate model 
are of next-to-leading importance.
We demonstrate that the original lattice data in \cite{DDM99} 
allow one to extract a reasonable value 
of the correlation scale $\lambda_q^2$ 
being in agreement with our vacuum condensate models.
At the very end, the lattice data support our conclusion 
on the shape of the pion distribution amplitude~\cite{MR86,MR92,BM98},
and vice versa, the CLEO experimental data~\cite{CLEO98} 
on pion photoproduction
confirm our suggestion about the range of the correlation scale values 
in an independent way~\cite{BMS01}.

%%%%%%%%%%%%%%%%%%%%%%%%%%%%%%%%%%%%%%%%%%%%%%%%%%%%%%%%%%%%%%%%%%%
\section{Models of nonlocal quark condensates}
 \label{NLC}
%%%%%%%%%%%%%%%%%%%%%%%%%%%%%%%%%%%%%%%%%%%%%%%%%%%%%%%%%%%%%%%%%%%
 \textbf{Small $x^2$ behavior.}
Let us start with the general properties 
of gauge-invariant nonlocal quark condensates (NLCs)
$F_{S,V}(x^2)$ 
following from their definitions 
\begin{eqnarray}
 \va{:\!\q_{\si}(0)E(0,x)q_{\ro}(x)\!:}
  = \frac{\qq{}}{4} 
    \left[F_S(x^2)-\frac{i\hat x}{4}F_V(x^2)
    \right]_{\ro \si},\quad
 E(0,x)
  = P\exp\left(-i\int\limits_0^x A_\mu(y)dy^\mu\right),
 \label{eq:NonQ} 
\end{eqnarray} 
where $\si,~\ro$ are spinor indices, 
the integral in the Fock--Schwinger string $E(0,x)$ is taken along
a straight-line path.
First of all, the condensates $F_{S,V}(x^2), \ldots$ 
should be analytic functions around the origin, 
and their derivatives at zero are related 
to condensates of the corresponding dimension.
Expanding $F_{S,V}(x^2)$ in the Taylor series 
at the origin in the fixed-point gauge $A_\mu(y)y^{\mu}=0$ 
(hence $E(0,x)=1$), 
one can obtain ~\cite{Gro94}:
\begin{eqnarray}
 \label{eq:QS}
  F_S(x^2)
  &=& \frac{\va{\q(0)q(x)}}{\qq{}}
   =  \frac{\qq{\left(1 + \frac1{2!}(x D)^2
                        + \frac1{4!}(x D)^4
                        + \cdots
                \right)}}{\qq{}}
 \nonumber\\
 &=& 1 + \left(\frac{Q^5}{2!2Q^3}-\frac{m_q^2}{2}
         \right)
         \left(\frac{x^2}4\right)
       + 2 \frac{Q^7_{(0)} - m_qQ^6_{(m_q)}}{4!3Q^3}
           \left(\frac{x^2}4\right)^2
       + O(x^6)\,, \\
  F_V(x^2)
  &=& i \frac{4\va{\q(0)\hat{x}q(x)}}{\qq{}x^2}
   =  i \frac{4\qq{\hat{x}
               \left((x D)
                 + \frac1{3!}(x D)^3
                 + \frac1{5!}(x D)^5
                 + \cdots
               \right)}}{\qq{}x^2}
 \nonumber\\
 &=& m_q + \frac{Q^6 + m_q \left(3Q^5 - 6m_q^2Q^3
                           \right)}
                {3! 3Q^3}\left(\frac{x^2}4\right)
         + \frac{Q^8_{(0)} - m_qQ^7_{(m_q)}}{5! 3 Q^3}
            \left(\frac{x^2}4\right)^2
         + O(x^6)\,.
 \label{eq:QV}
\end{eqnarray}
Here 
$D_{\mu}= \partial_{\mu} -ig A^a_{\mu}t^a$ 
is a covariant derivative;
$m_q$, the quark mass; $J^a_{\mu}$, quark vector current;
and the expansion coefficients $Q^i$ 
appearing in (\ref{eq:QS})-(\ref{eq:QV}) 
are vacuum expectation values (VEVs) of quark-gluon operators $O_i$
of dimension $i$, $Q^i = \va{O_i}$. 
These expansions with the explicit expressions for the condensates $Q^i$ 
have been derived in~\cite{Gro94}, (see Appendix A).
Condensates of lowest dimensions 
\begin{equation}
 \label{eq:Q6} 
  Q^3 = \qq{}\,,\quad
  Q^5 = ig\qq{\left(G_{\mu\nu}\si_{\mu\nu}\right)}
    \equiv m_0^2 \cdot Q^3\,,\quad
  Q^6 = \va{t^aJ^a_\mu t^bJ^b_\mu}\,,
\end{equation}
form the basis of standard QCD SR~\cite{SVZ} and have been estimated,
while the higher-dimensional VEVs, 
$Q^7_{(0)}, Q^7_{(m_q)}, Q^8_{(0)} \ldots$ 
are yet unknown. 
Note here that the matrix element of the 4-quark condensate
$Q^6$ is not known independently. 
Instead, it is usually evaluated in the ``factorization approximation",
the accuracy is estimated to be about 20\%~\cite{SVZ}, 
\begin{equation} \label{four-q}
\qq{\Gamma_1 q \bar{q}\Gamma_2}
  \approx - \frac{\mbox{Tr}\left(\Gamma_1\Gamma_2\right)}{144}\qq{}^2.
\end{equation}
The ``mixed" condensate $Q^5$ is expressed in the chiral limit 
as $m_0^2\cdot Q^3$ or $2\lambda_q^2\cdot Q^3$
\begin{equation} \label{eq:lambda_q}
 \Ds 4\frac{d F_S(x^2)}{ d x^2}|_{x=0}
 = \frac{Q^5}{4Q^3}
 \equiv 
   \frac{m^2_0}{4}
 = \frac{\lambda_q^2}2\,,
\end{equation}
and the parameter $\Ds \frac{\lambda_q^2}2$ 
fixes the width of $F_S(x^2)$ around the origin.
%\medskip
This important quantity has been estimated within the QCD sum-rule approach 
\begin{equation} 
 \lambda_{q}^{2}
   = \Big\{\vphantom{\Big\}}
   \begin{array}{ll} 0.4 \pm 0.1\ \gev{2}& \mbox{\cite{BI82}}\,,\\
                     0.5 \pm 0.05\ \gev{2}& \mbox{\cite{piv91}}.
   \end{array}
 \label{eq:lambda}
\end{equation} 
Estimates of $\lambda_{q}^{2}$ from instanton approaches~\cite{DEM97}
are somewhat larger:
$\lambda_{q}^{2} \geq  2/\rho_{c}^{2}\approx 0.6~\gev{2}$
where $\rho_{c}\approx 1.7 - 2$~\gev{-1} 
is an average characteristic size of the instanton fluctuation 
in the QCD vacuum. Taking into account all these estimates, 
in the following, we put a rather wide window 
$0.6 \geq\lambda_{q}^{2} \geq 0.4~\gev{2}$
for its value (``QCD Range" in Figs.\ref{fig:GaussAnChir},~\ref{fig:AdvAnChir}).

\textbf{ Large $x^2$ asymptotics from HQET.}
The large-$|x|$ properties of the NLC $F_S(x^2)$ 
have been analyzed in detail in~\cite{Rad91} in the framework of QCD SRs
for heavy--quark effective theory (HQET) of heavy-light mesons.
It was demonstrated that for a large Euclidean $x$, 
NLC is dominated by the contribution of the lowest state 
of a heavy-light meson 
with energy $\Lambda = (M_Q - m_Q)|_{m_Q\to \infty}$, 
and $F_S(x^2)\sim \exp\{- |x| \Lambda \}$ 
(numerically, $\Lambda$ is around 0.45~$GeV$).
In the following, we shall take this asymptotic behavior 
for NLC at large $|x|$
\begin{equation}
 \label{eq:asympt}
  F_{S,~V}
   \sim
    e^{- |x| \Lambda_{S, V}}\,.
\end{equation}

\textbf{Hints from QCD SRs, Gaussian Ansatz.}
To relate the NLC behaviour with the properties of mesons via QCD SR, 
it is convenient to parameterize the $x^2$-dependence
of (\ref{eq:QS})-(\ref{eq:QV}) by the distribution functions 
$f_X(\alpha)$ \textit{a'la} $\alpha$-representation for a propagator
\begin{equation}
 \label{eq:fsv}
  F_{S,V}(x^2)
   = \int_{0}^{\infty} e^{-\alpha x^2/4}\,
      f_{S,V}(\alpha)\, d\alpha\,,
 \quad \text{where~}
 \int_{0}^{\infty}\!\!  f_{S,V}(\alpha)\,  d\alpha =
  \Big\{\vphantom{\Big\}}
   \begin{array}{ll}
       1,& \mbox{S-case};\\
       0, & \mbox{V-case,~chiral limit.}
             \end{array}
\end{equation}
Here we use the Euclidean interval $x^2 =- x_{E}^2>0$, 
and the subscript $E$ will be omitted below for simplicity.
The representation (\ref{eq:fsv}) allows one 
(i) to involve smoothly NLCs into diagrammatic techniques, and
(ii) to clarify the physical properties of NLCs.
Indeed, functions $f_{S,V,\ldots}(\alpha)$ introduced in~\cite{MR86}
describe the distribution of quarks over virtuality $\alpha$ 
in nonperturbative vacuum.
The moments of $f_{S,V,\ldots}$ coincide with 
Taylor expansion coefficients in (\ref{eq:QS})-(\ref{eq:QV}).
For example, we have in $S$ case in the chiral limit
\begin{eqnarray} \label{eq:norm}
 \int_{0}^{\infty} \alpha f_{S}(\alpha)\, d\alpha
  = \frac12
     \left(\frac{\va{:\!\bar q D^2 q\!:}}{\va{:\!\bar{q}q\!:}}
      \equiv 
       \lambda_q^2
     \right)\!,\quad
 \int_{0}^{\infty} \alpha^2 f_{S}(\alpha)\, d\alpha
  = \frac{Q^7_{(0)}}{18Q^3}\,, \ldots
\end{eqnarray}
with $\lambda_q^2$ meaning the average virtuality of vacuum quarks.
Higher moments of $f_S(\alpha)$ are connected with 
higher--dimensional VEVs.
The difference between the Ansatzes for $F_S(x^2)$ looks more pronounced 
just for its $f_S(\alpha)$-images. 
It is evident that distributions  extremely concentrated at the origin
$f_S(\alpha) \sim \delta(\alpha)$, $\delta^{(1)}(\alpha),\ \ldots$
correspond to separate terms of the Taylor expansion~(\ref{eq:QS})-(\ref{eq:QV}).
At the same time, 
$f_S(\alpha) = \textsc{const}$ simulates a free propagation with zero mass: 
$F(x^2)\sim \textsc{const}/x^2$. 
The Gaussian Ansatz
\begin{equation}
 \label{eq:Gauss}
  F_S^{G}(x^2)
  = \exp\left(-\frac{\lambda_q^2}8x^2 \right)\,,
\end{equation} 
takes account of a single scale -- ``inverse width" $\lambda_q$ 
 of the coordinate distribution 
and corresponds to the virtuality distribution 
$f_S = \delta(\alpha - \lambda_q^2/2)$.
It fixes only one main property of the nonperturbative vacuum --- 
quarks can flow through the vacuum with a nonzero momentum $k$,
and the average virtuality of vacuum quarks is $\va{k^2} = \lambda_q^2/2$.
The Gaussian behavior at very large $|x|$ 
does not correspond to the expected NLC asymptotics~(\ref{eq:asympt}).
But for the moment QCD SRs, that deal with the smearing quantities -- 
moments of distribution amplitudes~\cite{MR86,MR92},
form factors~\cite{MR90,BR91,GroY92},
the incorrect asymptotics of NLC 
as well as subtle details of the Ansatz shape
are expected to be not so important. 
It is interesting that the Gaussian behaviour is supported by a model 
of the nonperturbative propagator, \cite{Piv96}, based on simple ``local--duality"
arguments. Namely, Eq.(16) in \cite{Piv96} demonstrates a behaviour rather close 
to Eq.(\ref{eq:Gauss}) within a physically important region 1 Fm.
Unfortunately, corresponding $F_S(x^2)$ decays too quickly beyond this region,
it becomes negative and oscillating, which is physically unclear.

Certainly, a more realistic model of $f(\alpha)$ should possess a finite width:
we expect that it is a continuous function
concentrated around a certain value $\lambda_q^2/2$ 
and rapidly decaying to zero as $\alpha$ goes to $0$ or $\infty$.
Moreover, the continuous distribution $f_{S}(\alpha)$ over virtuality
is directly related to the pion distribution amplitude
$\varphi_{\pi}(x)$ (for details, see sect.4),
as it was demonstrated 
with the help of the ``nondiagonal" correlator in~\cite{Rad94,BM95}.
%%%%%%%%%%%%%%%%%%%%%%%%%%%%%%%%%%%%%%%%%%%%%%%%%%%%%%%%%%%%%%%%%%%%%%%%
%%%%%%%%%%%%%%%%%%%%%%%%%%%%%%%%%%%%%%%%%%%%%%%%%%%%%%%%%%%%%%%%%%%%%%%%
%%%%% FIGURE 0: Different Ansatze for NLCs                     %%%%%%%%%
%%%%%%%%%%%%%%%%%%%%%%%%%%%%%%%%%%%%%%%%%%%%%%%%%%%%%%%%%%%%%%%%%%%%%%%%
\vspace*{-1mm}
 \begin{figure}[hbt]
  \centerline{\includegraphics[width=0.45\textwidth]{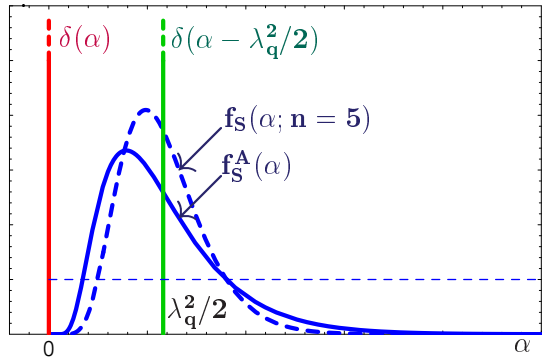}}
   \myfig{\label{fig:NLC}}
    \caption{\footnotesize 
   Graphical illustration of different NLC models 
   in $\alpha$-representation. 
   Thick bell-like (blue in a colour view) lines represents 
   the advanced Ansatz (solid line) in comparison with $f_S(\alpha;5)$
   (dashed line), 
   vertical line (green in a colour view) 
   at $\alpha=\lambda_q^2/2$  -- Gauss Ansatz,
   and vertical line (red in a colour view) 
   at $\alpha=0$  -- the local limit to standard condensates. 
   For comparison, we depict here the free propagation 
   $\alpha$-distribution in a dashed thin line.}
%  \end{minipage}
\end{figure}
%%%%%%%%%%%%%%%%%%%%%%%%%%%%%%%%%%%%%%%%%%%%%%%%%%%%%%%%%%%%%%%%%%%%%%%%

\textbf{``Advanced Ansatz".} 
To  construct  models of  nonlocal condensates,
one should satisfy some constraints. 
For instance, if we assume vacuum matrix elements
$\va{:\!\bar{q}(D^2)^{m}q\!:}$ to exist, 
then the function $f(\alpha)$ should decay faster than $1/\alpha^{m+1}$ 
as $\alpha \to \infty$.
If for all $m$, all such matrix elements exist, 
a possible choice could be a function
$ f(\alpha) \sim \alpha^n \exp(-\alpha \sigma)$, $etc.$ at large $\alpha$.
The opposite, small-$\alpha$, limit of $f(\alpha)$ 
is determined by the large-$|x|$ asymptotics~(\ref{eq:asympt})
of the function $F_S(x^2)$.
This means that
$f(\alpha) \sim \exp(-\Lambda^2/\alpha)$ in the small-$\alpha$ region.
By the simplest composition of both the asymptotics~\cite{Rad94,BM95},
we arrive at the class of Ansatzes~\cite{Rad94,BM95}
\begin{equation}
 f_S(\alpha;n) 
  \sim 
   \alpha^{n -1}\
    e^{-\Lambda^2/\alpha - \sigma^2 \alpha}
 \label{eq:A22}
\end{equation}
that gives a coordinate behavior
\begin{equation}
 \label{eq:A2}
  F_S(x^2;n)
   = {K_{n}\left(\Lambda \sqrt{4\sigma^2+|x^2|}\right)
      \over K_n\left(2\Lambda \sigma \right)}
       \left(\frac{2\sigma}{\sqrt{4\sigma^2 +|x^2|}}
       \right)^{n}.
\end{equation}
where $K_n(z)$ is the modified Bessel function.
The distribution $f_S^A(\alpha)\equiv f_S(\alpha;n=1)$ 
with the parameters $\Lambda_q \simeq 0.45~\mbox{GeV}$ 
and $\sigma_q^2 \simeq 10$~GeV$^{-2}$
is presented in Fig.1 in comparison with the distribution $f_S(\alpha;n=5)$.
In this case, the behavior of $F^{A}_S(x^2)\equiv F_S(x^2;1)$
is similar to that of the massive scalar propagator 
with a shifted argument:
\begin{equation}
 \label{eq:A2new}
  F^{A}_S(x^2)
  = {K_{1}\left(\Lambda \sqrt{4\sigma^2+|x^2|}
          \right)
     \over K_1\left(2\Lambda \sigma \right)}
      \frac{2\sigma}{\sqrt{4\sigma^2 +|x^2|}}\,,\quad
 \lambda_q^2 
  = 2 \frac{\Lambda}{\sigma}
    \cdot
     \frac{K_{2}\left(2\Lambda \sigma\right)}
          {K_1\left(2\Lambda\sigma\right)}\, . 
\end{equation}
The short-distance correlation scale $\lambda_q^2$ in (\ref{eq:A2new})
appears to be equal to $2/\sigma^2 $ at $\Lambda \sigma \ll 1$,
reproducing the single instanton--like result \cite{DEM97} 
where the parameter $\sigma$ imitates the instanton size $\rho_c$. 
In the opposite case at
$\Lambda \sigma \gg 1$, $\lambda_q^2$ is proportional to
$2\Lambda/\sigma$.
The asymptotics of $F^{A}_S(x^2)$ at large $|x| \gg 2\sigma$
\footnote{The asymptotics starts with $x^2 \gg 2$~Fm$^2$
at the mentioned value of $\sigma$-parameter.} 
is determined by the parameter $\Lambda$, 
\begin{equation}
 F^{A}_S(x^2)\mid_{x^2 \to \infty}
  \to \sigma \sqrt{\frac{2\pi}{\Lambda |x|^{3}}}\
   e^{-\Lambda x},
\end{equation}
where $\Lambda$ plays the role of an effective mass. 
This interpretation is transparent in the momentum representation of the NLC: 
$\tilde{F}^A_S(p^2)$ tends to the usual propagator form 
at small values of all the arguments
$$    
\tilde{F}^{A}_S(p^2) \sim {K_{1}\left( \sigma\sqrt{\Lambda^2+|p^2|}
          \right)
     \over K_1\left(2\Lambda \sigma \right)}
      \frac{\Lambda}{\sqrt{\Lambda^2 +|p^2|}}
 \rightarrow \frac{\Lambda^2}{\Lambda^2 + p^2} .
$$
``Advanced" Ansatz has been successfully applied 
in the nondiagonal QCD SR approach 
to the pion and its radial excitations~\cite{BM95}, 
and the main features of the pion have been described:
the mass spectrum of pion radial excitations $\pi'$ and $\pi''$
in close agreement with experimental data.

%%%%%%%%%%%%%%%%%%%%%%%%%%%%%%%%%%%%%%%%%%%%%%%%%%%%%%%%%%%%%%%%%%%
\section{Vacuum correlation length from lattice data}
 \label{lattice}
%%%%%%%%%%%%%%%%%%%%%%%%%%%%%%%%%%%%%%%%%%%%%%%%%%%%%%%%%%%%%%%%%%%
Here we consider the results of fitting the lattice simulation data 
for both the kinds of Ansatz $F_S^{G,A}(x^2)$ 
introduced previously for the nonperturbative part of a correlator.

\textbf{Processing the Pisa lattice data.}
The correlator $C_0(x)=- \sum_{f=1}^{N_f}\va{\Tr[\q^f(0)E(0,x)q^f(x)]}$
was measured on an Euclidean lattice~\cite{DDM99} at four points, 
$x=2a$, $4a$, $6a$, $8a$, inside 1 Fm, 
where $a\simeq 0.1$~Fm is the lattice spacing.
The simulation was performed with four flavors ($N_f=4$)
of staggered fermions with the same mass $m_q$ for every flavor 
and $SU(3)$ Wilson action for the pure gauge sector. 
Both the full QCD case  (\textbf{f}-QCD), \ie, 
including the effects of ``loop" fermions with the flavour masses 
$m_q a= 0.01,~0.02 $ 
and the quenched case (\textbf{q}-QCD
where the effects coming from loops of dynamical fermions are neglected)
with a set of flavour masses 
$m_q a= 0.01$, $0.02$, $0.05$, $0.10$ were considered, 
see Appendix B and~\cite{DDM99} for details.  
  
 The correlator can be written as a sum of a perturbative-like term,
$B \cdot F^{\text{pt}}_S(x^2)$,
proportional to $m_q/x^2$ at very short distances,
and a nonperturbative part, 
$A \cdot F^{\text{npt}}_S(x^2)$. 
Parameters of the latter are the main goal of the fit. 
For the perturbative part, the authors of~\cite{DDM99} 
have used just the short-distance asymptotics;
whereas for the nonperturbative one,
an exponential Ansatz 
$F^{\text{npt}}_S(x^2)=\exp(-\mu_0|x|)$
that corresponds to the asymptotics of very large distances
$x^2 \gg 1$~Fm$^2$.
In other words, 
these two approximations, involved in the fit simultaneously,
are adapted for different regions of $x$.
It seems more dangerously 
that the exponential Ansatz is a non-analytic function of $x^2$ 
at the origin:
it depends on $|x|=\sqrt{x^2}$ 
and its derivatives with respect to $x^2$ do not exist at the origin.
Therefore one loses any connection with local VEVs appearing in the OPE 
and with the corresponding interpretation of the correlation length, 
see section 2. 
Nevertheless, the fit of the lattice data has been performed~\cite{DDM99} 
and demonstrated very small $\chi^2$.
Note, the values of $\chi^2/N_{d.o.f.}$ in the Tables below
should be considered as purely indicative of the best fit quality.
One could not interpret these values in a standard statistical sense,
because their true norm is unknown, see brief discussion in ~\cite{DDM99}.
The extracted quantities, $A,~\mu_0,~B$\footnote{%
Note here that the authors of~\cite{DDM99} used
the dimensional coefficient $B_0$
related to our dimensionless parameter, $B_0=m_q B$.} 
are presented in Table 1 
(for every run) for a comparison with our results. 
\vskip 2mm

\centerline{\bf  Table 1: Modified\footnote{%
In the original table~\cite{DDM99}, the dimensionless combination 
$a B_0$ has been shown. 
Due to the previous footnote we recalculate 
the corresponding $B$ parameter for their fits.} 
fit~\cite{DDM99} ,~$\Ds F_S^{npt}(x)=\exp\left(- |x| \mu_0 \right)$}
\vskip 1mm

\moveright 13mm 
\vbox{\offinterlineskip
\halign{\strut
\vrule \hfil\quad $#$ \hfil \quad &
\vrule \hfil\quad $#$ \hfil \quad &
\vrule \hfil\quad $#$ \hfil \quad &
\vrule \hfil\quad $#$ \hfil \quad &
\vrule \hfil\quad $#$ \hfil \quad &
\vrule \hfil\quad $#$ \hfil \quad \vrule \cr
\noalign{\hrule}
\beta, ~{\rm theory} &
a \cdot m_q &
a^3 A_0 \times 10^2     &
a \mu_0     &
B  &
\chi^2/N_{d.o.f.}
%\cr & & & & & \cr \noalign{\hrule} \noalign{\hrule}
\cr \noalign{\hrule height 1pt}
5.35, ~{\rm \textbf{f}}  & 0.01  & 0.49(13)& 0.16(4) & 1.3(3) & 1.3\cdot 10^{-2} \cr
\noalign{\hrule}
5.35, ~{\rm \textbf{f}} & 0.02   & 1.7(5)  & 0.26(4) & 0.95(5)& 6.4\cdot 10^{-3} \cr
\noalign{\hrule}
6.00, ~{\rm \textbf{q}} & 0.01   & 1.6(5)  &0.16(4)  & 0.9(1) & 7.6\cdot 10^{-2} \cr
\noalign{\hrule}
5.91, ~{\rm \textbf{q}}  & 0.02  & 2.3(7)  & 0.26(3) & 1.25(7)& 5.2\cdot 10^{-2} \cr
\noalign{\hrule}
6.00, ~{\rm \textbf{q}} & 0.05   & 1.8(4)  & 0.34(2) & 1.4(2) & 0.2 \cr
\noalign{\hrule}
6.00, ~{\rm \textbf{q}} & 0.10   & 5.6(5)  & 0.55(1) &  \ 1.0(1)\, \ 
                                                              & 1.3 \cdot 10^{-2}
\cr\noalign{\hrule}
}}
\vskip 2mm

We test different NLC Ansatzes  and extract the parameters
of the correlator using three-step 
procedure.
At the first step, we fit rearranged formulas for the correlator.
Note that the masses of light flavors 
appear unnaturally large in the lattice simulation
and cannot be considered small 
(for the values of masses in MeV, have a look at the axes of Fig.3). 
For this reason, one should take account of all possible mass terms 
in both parts of $C_0$:
\begin{eqnarray}
&& C_0 (x)
  = B\cdot F^{\text{pt}}_S(x^2)
  + A\cdot F^{\text{npt}}_S(x^2),\\
 && B \sim \frac{N_f N_c}{\pi^2}; ~~F^{\text{pt}}_S(x^2)
   =  m_q \left(\frac{m_q}{x} K_1(m_q x)\right);\quad
  F^{\text{npt}}_S(x^2)
   =  F^{G,A}_S(x^2) + \frac1{8}m_q^2x^2 +\ldots\,.
 \label{eq:basic}
\end{eqnarray}

Following~\cite{DEJM00}, we fix the perturbative-like part 
in the form of free propagation 
$F^{\text{pt}} = \frac{m_q}{x} K_1(m_q x)$ 
with mass $m_q$.
For nonperturbative part, we keep all the known mass-terms 
from the expansion (\ref{eq:QS}),
see the expression for $Q^6_{(m_q)}$ in Appendix~A.
\vskip 1mm

\centerline{\bf Table 2: Fit for the Gaussian Ansatz,
$\Ds F_S^G(x)=\exp\left(- {x^2 \lambda_L^2 \over 8}\right)
$}
\vskip 1mm

\moveright 13mm 
\vbox{\offinterlineskip 
\halign{\strut
 \vrule\hfil\quad $#$ \hfil\quad&
 \vrule\hfil\quad $#$ \hfil\quad&
 \vrule\hfil\quad $#$ \hfil\quad&
 \vrule\hfil\quad $#$ \hfil\quad&
 \vrule\hfil\quad $#$ \hfil\quad&
 \vrule\hfil\quad $#$ \hfil\quad
 \vrule\cr
\noalign{\hrule}
 \beta, ~{\rm theory}
      & a \cdot m_q
             & a^3 A_0 \times 10^2
                         & (a\cdot \lambda_L)^2
                                     & B 
                                               & \chi^2/N_{d.o.f.}
\cr \noalign{\hrule height 1pt}
 5.35, ~{\rm \textbf{f}}
      & 0.01 & 0.28(4)   & 0.10(3)   & 1.67(14)& 0.075  \cr \noalign{\hrule}
 5.35, ~{\rm \textbf{f}}
      & 0.02 & 0.69(12)  & 0.18(3)   & 1.67(21)& 0.125  \cr \noalign{\hrule}
 6.00, ~{\rm \textbf{q}}
      & 0.01 & 0.93(14)  & 0.10(3)   & 2.19(65)& 0.01   \cr \noalign{\hrule}
 5.91, ~{\rm \textbf{q}}          
      & 0.02 & 0.92(14)  & 0.16(3)   & 2.31(27)& 0.65   \cr \noalign{\hrule}
 6.00, ~{\rm \textbf{q}}
      & 0.05 & 0.60(7)   & 0.23(2)   & 1.7(6)  & 1.78   \cr \noalign{\hrule}
 6.00, ~{\rm \textbf{q}}
      & 0.10 &  -       & -        &  -     & 2.7    \cr \noalign{\hrule}
}}
\vskip 2mm

\centerline{\bf Table 3: Fit for the ``Advanced" Ansatz, $F^{A}_S(x), 
\Lambda = 0.5$~GeV}
\vskip 1mm

\moveright 13mm 
\vbox{\offinterlineskip
\halign{\strut
 \vrule\hfil\quad $#$ \hfil\quad&
 \vrule\hfil\quad $#$ \hfil\quad&
 \vrule\hfil\quad $#$ \hfil\quad&
 \vrule\hfil\quad $#$ \hfil\quad&
 \vrule\hfil\quad $#$ \hfil\quad&
 \vrule\hfil\quad $#$ \hfil\quad
 \vrule \cr
\noalign{\hrule}
 \beta, ~{\rm theory}
      & a \cdot m_q
             & a^3 A_0 \times 10^2
                         &(a\cdot \lambda_L)^2
                                     & B      & \chi^2/N_{d.o.f.}
\cr \noalign{\hrule height 1pt}
 5.35, ~{\rm \textbf{f}}
      & 0.01 & 0.29(5)   & 0.11(3)   &1.63(19)&0.03  \cr \noalign{\hrule}
 5.35, ~{\rm \textbf{f}}          
      & 0.02 & 0.76(16)  & 0.23(5)   &1.57(30)&0.05  \cr \noalign{\hrule}
 6.00, ~{\rm \textbf{q}}          
      & 0.01 & 0.97(17)  & 0.12(3)   &1.95(63)&5\cdot 10^{-4}  \cr \noalign{\hrule}
 5.91, ~{\rm \textbf{q}}          
      & 0.02 & 1.04(20)  & 0.21(5)   &2.12(40)&0.3   \cr \noalign{\hrule}
 6.00, ~{\rm \textbf{q}}          
      & 0.05 & 0.79(14)  & 0.38(6)   &1.59(9) &0.6    \cr \noalign{\hrule}
 6.00, ~{\rm \textbf{q}}          
      & 0.10 & 2.72(32)  & 1.41(7)   &1.13(6) &0.3   \cr \noalign{\hrule}
      }}
\vskip 2mm

\begin{figure}[thb]
  \centerline{\includegraphics[width=\textwidth]{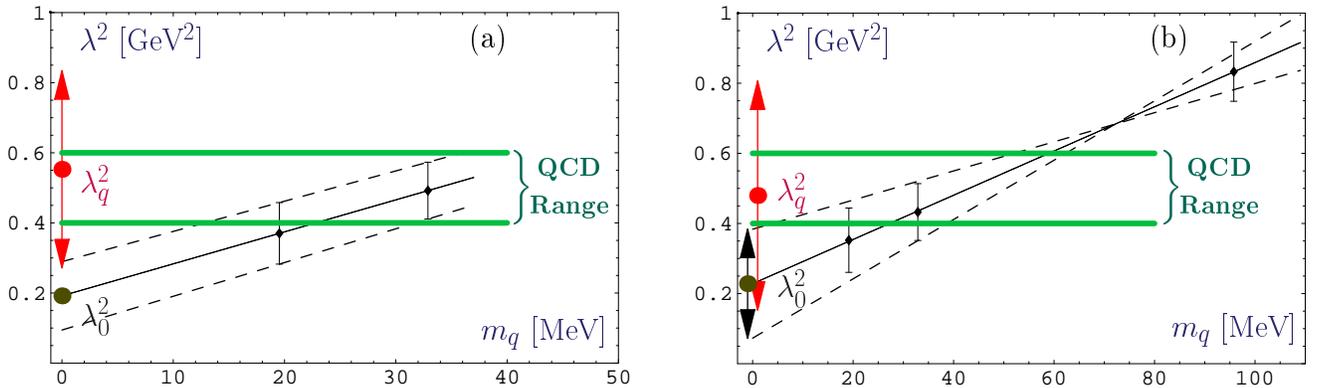}}
   \vspace*{1mm}\myfig{\label{fig:GaussAnChir}}
    \caption{\footnotesize
   Graphical illustration of the fit results for $\lambda_L^2$ 
   -- black points with error bars; 
   for the chiral limit 
   ( $m_q \to 0$) 
   and the evolution to the scale $1$~GeV$^2$ 
   for the Gaussian Ansatz both in 
   full (part (a)) and 
   quenched (part (b)) LQCD.
   Dashed black lines on both parts correspond to chiral limit
   procedure with taking into account error-bars,
   black blobs -- to the central points of the chiral limit,
   and red blobs -- to the resulting values of $\lambda_q^2$ on the
   QCD scale of $1$~GeV$^2$. 
   Red arrows show the overall error-bars of the extracted 
   $\lambda_q^2(\mu^2=1~\gev{2})$, 
   whereas green thick lines bound the QCD preferred region.}
\end{figure}

As a result of the fit, we extract, from lattice data,
an {\it intermediate} correlation scale  $ \lambda_L$ %(``inverse width")
that parameterizes $F_S^{G,A}(x^2)$ in (\ref{eq:basic}) 
instead of $\lambda_q$,  
$$\lambda^2_L = -8 \frac{d F_S^{G,A}(x^2)}{d x^2}|_{x = 0},$$
and depends on lattice conditions,
namely $m_q$, $a$, and $\beta$. 
We expect that this quantity coincides in the chiral limit  
with $\lambda_q^2$, determined in (\ref{eq:lambda_q}) within the massless QCD,
\ie, 
$\lambda_L^2(m_q)\mid_{m_q \to 0} \longrightarrow \lambda_0^2
= \lambda_q^2(\mu_L^2)$
on the lattice normalization scale $\mu_L^2$.
The results for $A,~B$ and $\lambda_L^2(m_q)$, 
obtained in various cases\footnote{%
The Gaussian Ansatz has been tested 
in a lattice in~\cite{Meg99} 
(after our suggestion in a private communication)
 without any mass corrections. 
The fit-result, $\lambda_L^2(m_q \cdot a =0.01)=0.46(5)$, 
appears somewhat larger than our fit-result in Fig.\ref{fig:GaussAnChir}(a).}
are collected in 
Tables~2 and 3;
let us outline the main features:

(i) The $\chi^2$ for the  cases of both Ansatzes     look higher 
    than in Table 1, but still sufficiently small, 
    especially for \textbf{f}-lattice.
    An exception is the last run for the \textbf{q}-lattice with the largest mass 
    $m_q \cdot a=0.10$ 
    that corresponds to $m_q \simeq 200$~MeV~\cite{DDM99}.
    To process the data 
    with so a huge mass, one should involve a lot of mass-terms 
    into the fit formula (\ref{eq:basic}). 
    For this reason, we exclude this run results from subsequent analysis.
     
(ii) The extracted coefficient B that fixes the perturbative-like contribution 
     should not change significantly from one run to another
     for both Ansatzes  and for both kinds of lattices.
     This property can signal about a good quality of the fits, 
     it confirms the reliability of the fit at least for the \textbf{f}-lattice case, 
     see Tables 1 and 2.
     
(iii) The fitted parameter $\lambda_L^2(m_q)$ 
      that fixes the behavior of the nonperturbative part
      has a strong and monotonic dependence on $m_q$, 
      in contrast with the parameter $B$, see item~(ii).

\begin{figure}[h]
  \centerline{\includegraphics[width=\textwidth]{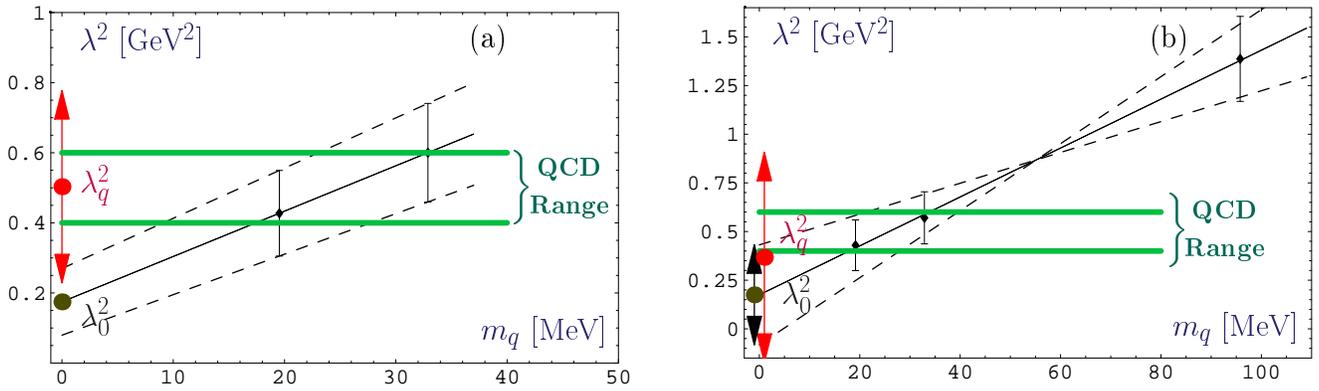}}
   \vspace*{1mm}\myfig{\label{fig:AdvAnChir}}
    \caption{\footnotesize
    Graphical illustration to three-step procedure
    for the case of Advanced Ansatz  both in 
   full (part (a)) and  quenched (part (b)) LQCD.
    } 
\end{figure}

%%%%%%%%%%%%%%%%%Extrapolation%%%%%%%%%%%%%%%%%%%%%%%%%%%%%%%%%%%%%%%%%%%%%%%%%%%%%%%%
\textbf{ Extrapolation to the chiral limit.}
At the second step, 
we extrapolate the intermediate $\lambda_L^2(m_q)$ to the chiral limit, 
suggesting a simple linear dependence on $m_q \to 0$. 
The linear law seems to be rather naive, 
but the observed strong dependence of $\lambda_L^2(m_q)$ 
on the quark mass %(see Table1,~2)
is well supported by the data, 
see graphics in Fig.\ref{fig:GaussAnChir},~\ref{fig:AdvAnChir}. 
Really, the linear extrapolation of the first three run results 
for the \textbf{q}-lattice 
is self-consistent for both Ansatzes
(Fig.\ref{fig:GaussAnChir}(b),~\ref{fig:AdvAnChir}(b)),
and the results are in a reasonable agreement with the 
corresponding \textbf{f}-lattice results
(Fig.\ref{fig:GaussAnChir}(a),~\ref{fig:AdvAnChir}(a)).

%%%%%%%%%%%%%%%%%%%%%%%%%%%%%%%%%%%%%%%%%%%%%%%%%%%%%%%%%%%%%%%%%%%%%%%%
%%%%% FIGURE 1g: The evolution+chiral limit for Gauss Ansatz   %%%%%%%%%
%%%%%%%%%%%%%%%%%%%%%%%%%%%%%%%%%%%%%%%%%%%%%%%%%%%%%%%%%%%%%%%%%%%%%%%%
\vspace*{1mm}
%%%%%%%%%%%%%%%%%%%%%%%%%%%%%%%%%%%%%%%%%%%%%%%%%%%%%%%%%%%%%%%%%%%%%%%%
%%%%% FIGURE 3: The evolution+chiral limit for Adv.Ansatz   %%%%%%%%%%%
%%%%%%%%%%%%%%%%%%%%%%%%%%%%%%%%%%%%%%%%%%%%%%%%%%%%%%%%%%%%%%%%%%%%%%%%
\vspace*{1mm}
%%%%%%%%%%%%%%%%%%%%%%%%%%%%%%%%%%%%%%%%%%%%%%%%%%%%%%%%%%%%%%%%%%%%%%%%

The explicit results for $\lambda_0^2$ 
are presented in the first line of Table 4:% for details:
~note, $\lambda_0^2$ for the \textbf{q}-lattice is not smaller 
than for the \textbf{f}-one for both Ansatzes; 
the large error bars for $\lambda_0^2$
appear due to the roughness of the chiral extrapolation procedure.

To clarify the reliability of the approximation, 
we repeat the same extrapolation procedure 
with the dimensionless ``lattice quantities", 
$L^2=(a\cdot \lambda_L)^2$ and $M=a\cdot m_q$. 
Lattice spacing $a$ involved into this extrapolation
is different for different runs 
(see Appendix B, Eqs.(\ref{ref:f})-(\ref{ref:q})). 
To return to  ``physical" quantity $\lambda_0$ at the very end, 
we adapt an average spacing $a_q$ 
corresponding 
to the \textbf{q}-lattice case, 
and a similar average one $a_f$ -- to the \textbf{f}-lattice case 
(see Appendix~B). 
It is clear, that this procedure is even more crude 
than the previous one. 
Nevertheless, the corresponding $\lambda_0$ 
well agree 
with the previous ``three-point" \textbf{q}-lattice result,
compare the first and third lines in Table 4.
But the procedure falls down for \textbf{f}-lattice data, 
the results appear too small. 

At the third step, we perform 
\textbf{an evolution to continuum normalization scale}.
To compare the %lattice 
results for the shape of NLC $F_S(x^2,\mu^2)$
(or distribution $f(\alpha, \mu^2)$ at scale $\mu^2$) 
with the same quantity
taken on another scale $Q^2$,
the corresponding evolution law in $\mu^2$ is required.
For a general case it looks as ambitious and rather complicated problem
that is yet unsolved.
But in our partial case 
if we fix how $\lambda_q^2(\mu^2)$ evolves with $\mu^2$,
then the evolution law for both Ansatze is also fixed.
Therefore, we shall consider 
the evolution of single characteristic of the shape -- $\lambda_q^2(\mu^2)$,
setting the equivalence $\lambda_q^2(\mu_L^2) = \lambda_0^2$ in a lattice.
We live not in a lattice, 
but, instead, in a continuum.
Therefore we need a procedure 
to relate
lattice measurements with our observables in continuum, 
the corresponding one-loop evolution law 
for this kind of a lattice presented, \eg, 
in~\cite{Alt93},
\begin{equation} 
 \label{eq:evolut}
  \Ds \lambda_q^2(\mu^2)\cdot
   \left[\ln(\mu/\Lambda_{MS})
   \right]^{\Ds \frac{\gamma_{\lambda}}{b_0}}
 =
  \lambda_0^2 \cdot
   \left[\frac{4\pi^2}{b_0}\beta
        +\ln(\Lambda_L/\Lambda_{MS})
        +6.54
  \right]^{\Ds \frac{\gamma_{\lambda}}{b_0}}\,.
\end{equation}
The continuum quantities enter into the l.h.s. of Eq.(\ref{eq:evolut});
while the lattice quantities, into the r.h.s..
Here $\Lambda_{MS}/\Lambda_L =76.44$, 
$b_0=11/3C_A-4/3T_R N_f$ -- first coefficient 
of the $\beta$-function, 
and $\gamma_{\lambda} = (2C_A-8C_F) = - 14/3$ -- 
one-loop anomalous dimension 
for $\lambda_q^2(\mu^2)$, calculated in ~\cite{Mor84}. 
So, using formula (\ref{eq:evolut}) 
to return  to continuum 
$\lambda_q^2(\mu^2=1\ \gev{2})$, 
we obtain the final results for $\lambda_q^2$
appearing just in the ``QCD range",
see Figs.\ref{fig:GaussAnChir},~\ref{fig:AdvAnChir}. 
Let us consider these evolved results, 
presented in Table 4, in more detail. 
%\noindent

(i)~The Gaussian Ansatz. 
The mean values of $\lambda_q^2$ for \textbf{f}- and \textbf{q}-lattices 
become closer to one another after the evolution 
starting from the corresponding different values of $\lambda_0^2$. 
This ``focusing" effect is due 
to a difference of the evolution in both the cases: 
for the \textbf{f}-lattice ($\beta = 5.35; ~b_0 = 25/3, N_f=4$)
the transition factor from Eq.(\ref{eq:evolut}) is $2.88$, 
while for \textbf{q}-lattice ($\beta \approx 6;~b_0 = 11, N_f=0~$)
-- $2.11$.
Note here that if one attempts to exchange, by hand, 
these evolution laws 
(\textbf{q} for \textbf{f} and vice versa),
then the final numbers diverge out of the QCD range.
The observed focusing may demonstrate a complementarity 
of the chiral limit results 
to the evolution law.

(ii)~The Advanced Ansatz. 
The value of $\lambda_q^2$
for \textbf{f}-lattice, $0.5 ~\gev{2}$, 
appears to be very close to an average $\lambda_q^2$ 
for the Gaussian Ansatz in the left side of Table~4. 
The result for the \textbf{q}-lattice 
is less than this average estimate and located 
near the low boundary of the QCD range. 
But, in virtue of huge error bars,
the latter result does not contradict the \textbf{f}-lattice result.

Finally we can conclude
that Pisa lattice data reported in~\cite{DDM99} really ``feel" 
the short distance correlation scale in the QCD vacuum. 
The data processing 
explicitly demonstrates that the extracted mean values 
of $\lambda_q^2$ 
are in agreement with the estimates 
from the QCD SR approach (\ref{eq:lambda})
for all the considered cases.
Moreover, the results are in agreement 
with the old lattice result, 
$\lambda_q^2 \approx 0.55$~\gev{2}, 
obtained in~\cite{KS87} on a \textbf{q}-lattice.
Huge errors bar are the problem of these estimates, 
and this does not allow one to confirm the agreement with the ``QCD range" 
once and for all. 
The main uncertainty follows from the chiral limit procedure 
(``second step").
To reduce the uncertainty, one should improve 
the theoretical part in (\ref{eq:basic}) 
as well as the ``lattice" part of the fit. 
Namely, we need more numbers of the lattice runs 
with a moderate/small quark masses $M=a\cdot m_q$ 
for a reliable extrapolation; 
to process all existing results, 
we should include the most important subset of mass-terms 
into the nonperturbative part of the fitted formula~(\ref{eq:basic}).
%%%%%%%%%%%%%%%%%%%%%%%%%%%%%%%%%%%%%%%%%%%%%%%%%%%%%%%%%%%%%%%%%%%%%%%
%%                 T A B L E   4                                       %
%%%%%%%%%%%%%%%%%%%%%%%%%%%%%%%%%%%%%%%%%%%%%%%%%%%%%%%%%%%%%%%%%%%%%%%%
\vskip 1mm

\centerline{%
 \textbf{Table 4: On the analysis of different approaches 
         to the chiral limit
         }
          }
   \vskip 2mm

\noindent\hspace*{0.03\textwidth}
\begin{minipage}{0.94\textwidth}
\begin{tabular}{||c||c|c||c|c||}\hline
%___________________________________________________________________
 &\hspace{0.1mm} Full  QCD $\vphantom{^{\big|}_{\big|}}$ \hspace{0.1mm}
   &\hspace{0.1mm} Quenched  QCD \hspace{0.1mm}
     &\hspace{0.1mm} Full  QCD \hspace{0.1mm}
       &\hspace{0.1mm} Quenched QCD \hspace{0.1mm}
\\ 
 \cline{2-5}
 &\multicolumn{2}{c||}
  {\strut\vphantom{$^\big|_\big|$}\hspace{0.1mm}
   \hspace{0.1mm}$\lambda_q^2$~[GeV$^2$]\hspace{0.1mm},
    Gauss-Ansatz \hspace{0.1mm}}
     &\multicolumn{2}{c||}
  {\strut\hspace{0.1mm}
    \hspace{0.1mm}$\lambda_q^2$~[GeV$^2$]\hspace{0.1mm},
     \hspace{0.1mm}Advanced-Ansatz\hspace{0.1mm}} \\
 \hline\hline
%___________dimensional_____________________________________________
\parbox{33mm}{$\vphantom{^{\big|}}$ 
  Dimensional\\ Chiral Limit $(\lambda_0^2)\vphantom{_{\big|}}$}
 &$0.19\pm 0.10$
   &$0.23 \pm 0.16$
     &$0.175 \pm 0.1$
       &$0.175 \pm 0.26$ \\ \hline
%___________dimensional+evolution____________________________________
\parbox{23mm}{$\vphantom{^{\big|}}$
  Evolution to \\ $\mu_0^2=1$ GeV$^2\vphantom{_{\big|}}$}
 &${\bf 0.55\pm 0.28}$  
  &${\bf 0.48 \pm 0.33}$
     &${\bf 0.50 \pm 0.27}$
       &${\bf 0.37\pm0.54}$ \\ \hline\hline
%___________adimensional_____________________________________________
\parbox{33mm}{$\vphantom{^{\big|}}$ Dimensionless\\
   Chiral Limit $(\lambda_0^2)\vphantom{_{\big|}}$}
 &$ \times$
   &$0.24\pm0.15$
     &$ \times $
       &$0.19 \pm0.23 $ \\ \hline
%___________adimensional+evolution___________________________________
\parbox{23mm}{$\vphantom{^{\big|}}$
  Evolution to \\ $\mu_0^2=1$ GeV$^2\vphantom{_{\big|}}$}
 &$ \times$
   &${\bf 0.50\pm0.31}$
     &$ \times$
       &${\bf 0.40 \pm 0.50}$ \\  \hline
%____________________________________________________________________
\end{tabular}\\[3mm]

\noindent Results of extrapolation to  $m_q \to 0$ 
are evolved to the standard normalization scale $\mu_0^2 = 1$~GeV$^2$.
Crosses in full QCD columns mean the breakdown of chiral limit procedure.
\end{minipage}
\vskip 3mm

Another problem is to extract the quark condensate value $|\qq{}|$
from the lattice QCD data. 
In the fit it is just the coefficient $A$ divided by 
the number of flavors $N_f=4$. 
But the three-step procedure fails for this data\footnote{%
The authors of \cite{DDM99} also did not obtain 
reasonable estimates for $\qq{}$ using this kind of data,
therefore they have performed an individual measurement of the quantity,
see Eq.(3.3-3.5) in \cite{DDM99}}
providing too small values for the quark condensate.  

Another possibility is
to construct the RG-invariant  quantity $m_q(\mu^2)\qq{}(\mu^2)$ 
in the lattice 
to avoid  both the chiral limit and the renormalization effects. 
In this way, 
we obtained different estimates for every run 
in full the lattice QCD; the estimate region is
\begin{equation}
 \label{eq-QCRGI}
 |2 m_q \qq{}| =
  (2 - 5)\cdot 10^{-4}\ \gev{4}\ ,
\end{equation}  
that should be compared with the well-known value
fixed by the current algebra
\begin{eqnarray}
 \label{eq:curr-algebra}
&&   |\va{m_u \bar{u}u} + \va{m_d \bar{d}d}|
  \ =\ {1 \over 2}f_{\pi}^2 m_{\pi}^2(1+O(m_{\pi}^2))
  \ \approx\  1.7\cdot 10^{-4}~\gev{4}.
\end{eqnarray}
The estimate (\ref{eq-QCRGI}) produces for real QCD case 
with the current masses $m_{u,d}\simeq 5.5$~MeV
\begin{equation}
 \label{eq-QC1GeV}
  \left|\qq{}(1\ \gev{2})\right|^{1/3} =
  265 - 358\ \Mev 
  ~~\text{vs the standard value}~250\ \Mev,~\cite{SVZ}.
\end{equation}  
The values are overestimated, although the lowest one
corresponding to the run with $m_q \cdot a =0.01$
looks reasonable.

\section{Nonlocal quark condensates and pion distribution \\
amplitude}
The pion distribution amplitude (DA) of twist-2,
$\varphi_\pi(x,\mu^2)$,
is a gauge- and process-independent characteristic of the pion
that  universally specifies the longitudinal momentum $xP$
distribution of valence quarks in the pion with momentum $P$
(see, e.g., \cite{CZ84} for a review),
\begin{equation}
 \va{0\mid\bar d(0)\gamma^\mu\gamma_5 E(0,z)u(z)\mid \pi(P)}\Big|_{z^2=0}
  = i f_{\pi}P^{\mu}
     \int^1_0 dx e^{ix(zP)}\ \varphi_{\pi}(x,\mu^2)\ .
\end{equation}
Due to factorization theorems \cite{CZ77,ERBL79}, 
it enters as the central input of
various QCD calculations of hard exclusive
processes.
Here we illustrate how a value of the correlation scale
$\lambda_q^2$ ($\sim 1/l^2$) 
can affect the shape of the pion DA.
First, we consider NLC QCD SR for DA moments
that provides the smearing quantities, %$\va{\xi^N}$,
moments 
$\Ds \va{\xi^N}_{\pi}= \int_0^1 (2x-1)^N \varphi_{\pi}(x)dx $,
to restore a profile $\varphi_{\pi}(x)$ of the pion DA.
The NLC QCD SR are based on different kinds of
Gaussian Ansatzes \cite{MR86,MR92} for NLCs
that naturally appear in the theoretical (r.h.s.) part of the SR.

\textbf{Pion DA from NLC QCD SR for pion DA moments. The Gaussian Ansatz.}
The SR involves 5 different kinds of nonlocal condensates
in addition to the scalar condensate contribution, 
$\Delta\Phi_S(x;M^2)$, 
for details see \cite{MR86,MR92,BM98}. 
%%%%%%%%%%%%%%%%%%%%%%%%%%%%%%%%%%%%%%%%%%%%%%%%%%%%%%%%%%%%%%%%%%%%%%%%
%%%%% FIGURE 2: The set of admissible models for the pion DA %%%%%%%%%%%
%%%%%%%%%%%%%%%%%%%%%%%%%%%%%%%%%%%%%%%%%%%%%%%%%%%%%%%%%%%%%%%%%%%%%%%%
\vspace*{1mm}
 \begin{figure}[hb]
  \centerline{\includegraphics[width=\textwidth]{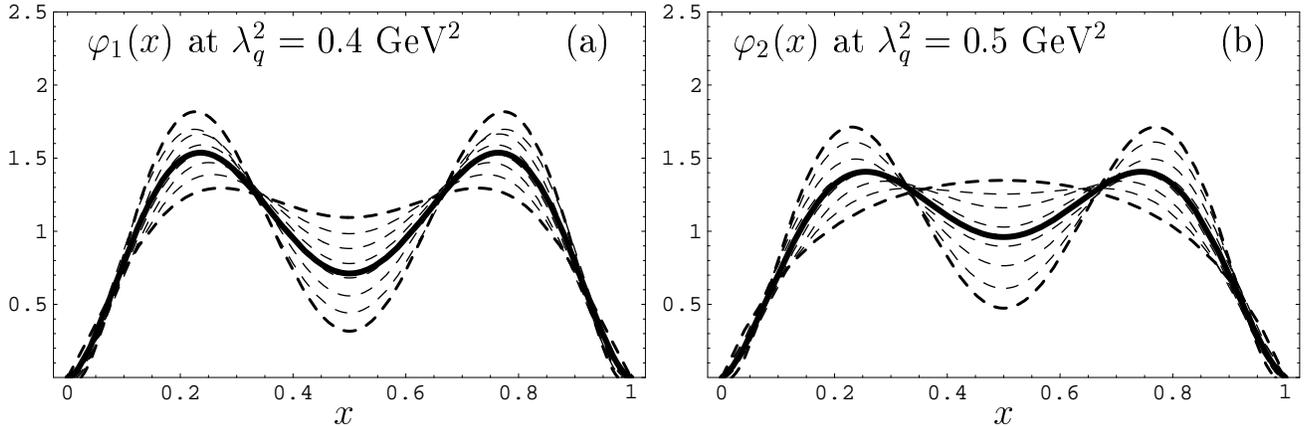}}
   \vspace*{1mm}\myfig{\label{fig:DArange}}
    \caption{\footnotesize
     Graphical representation of $\varphi_{\pi1}(x,\mu^2)$ (part (a)) 
     and $\varphi_{\pi2}(x,\mu^2)$ (part (b)) 
     at the characteristic renormalization point
     $\mu^2 \simeq 1~\gev{2}$.
     The thick solid lines in both plots denote
     $\varphi^{opt}_{\pi1,2}(x)$, i.e., 
     the best fit to the determined values of the moments, 
     whereas dashed lines illustrate admissible options
     approximately with $\chi^2\ \leq 1$.}
 \end{figure}
%%%%%%%%%%%%%%%%%%%%%%%%%%%%%%%%%%%%%%%%%%%%%%%%%%%%%%%%%%%%%%%%%%%%%%%%
The scalar NLC contribution results from the
``factorization approximation" to the nonlocal four-quarks condensate,
and its accuracy is yet unknown, compare with the approximation, Eq.(\ref{four-q}).
But in any case the contribution $\Delta\Phi_S(x;M^2)$ is numerically the largest one
for not too high moments.
Therefore, the main features of the shape of $\varphi_{\pi}(x)$ in 
Fig.\ref{fig:DArange} is, roughly speaking, 
the net result of the interplay 
between the perturbative contribution and the nonperturbative term
$\Delta\Phi_S(x;M^2)$ that dominates the r.h.s. of the SR. 

The QCD SR predicts the values of moments $\va{\xi^N}_{\pi}(\mu^2\simeq 1~\gev{2}$) 
within their error bars. 
For this reason, one obtains, after restoration a ``bunch" of admissible 
DA profiles \cite{BMS01}
corresponding to the moment error bars, 
rather than a single sample of profile.
These profiles are shown in Fig.\ref{fig:DArange} 
by dashed lines, 
in addition to the optimal one (thick solid line) 
that corresponds to the best fit 
at $\chi_{opt}^2\ \sim 10^{-3}$.
Comparing DA profiles at different values of $\lambda_q^2$ 
in Fig.\ref{fig:DArange}(a) and (b), one can conclude 
that the larger is the correlation scale the smaller is the concavity 
in the middle of the profiles, 
and their shape becomes closer to the shape of ``asymptotic" DA 
$\varphi_{as}(x)=6x(1-x)$. Therefore, a trial bunch (is not shown here) 
corresponding to 
the value $\lambda_q^2= 0.6~\gev{2}$ at the boundary of the introduced QCD range
contains mainly convex in the mid-point profiles that are close 
to the asymptotic one.

\begin{figure}[ht]
 \myfig{\label{fig:ogurec}}
  \centerline{\includegraphics[width=0.7\textwidth]{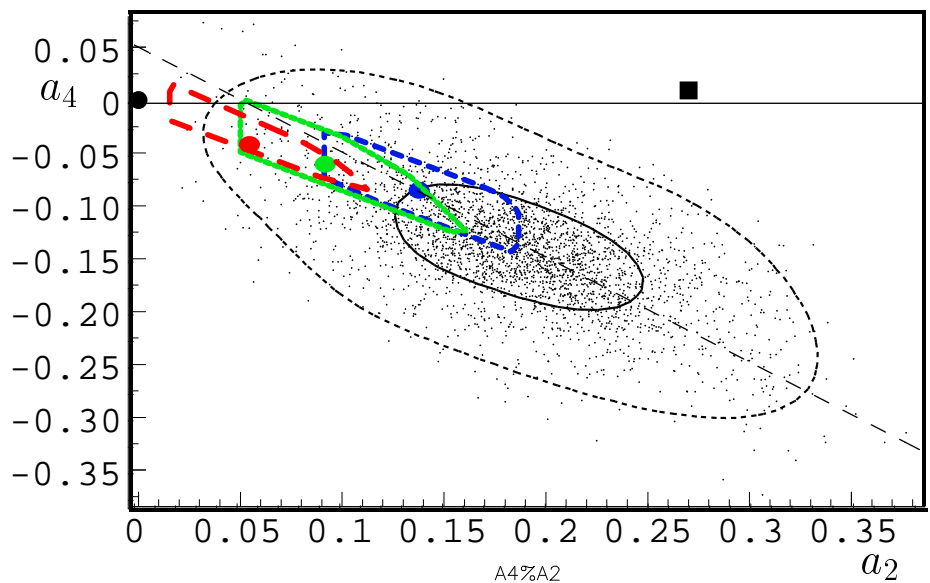}}
   \caption{\footnotesize    
   The parameter space of ($a_{2}$, $a_{4}$) pairs,
   corresponding to the allowed values of the second
   and fourth Gegenbauer coefficients: (i) calculated within the NLC-SR
   approach for three different values of
   $\lambda_{q}^{2}$ and evolved to $\mu^2_{S\&Y}=~5.6 ~\gev{2}$;
   (ii) confidence regions extracted by Schmedding and Yakovlev 
   \protect\cite{SchmYa99} from CLEO data \protect\cite{CLEO98}.
   Contour lines show $68\%$ (solid line) and $95\%$ (dashed line)
   confidential regions.
   It is shown how our estimate of the confidence region with
   $\lambda_{q}^{2}=0.4,~0.5,~0.6 ~\gev{2}$  overlaps
   with those displayed in Fig.~6 of Ref.\protect\cite{SchmYa99}.
   Bold dot in the plot marks the parameter pairs for the asymptotic DA,
   full square -- for Chernyak--Zhitnitsky DA.}            
\end{figure}

We have established in~\cite{BMS01} 
that a two-parameter model $\varphi_{\pi}(x;a_2, a_4)$, 
the parameters being the Gegenbauer coefficients $a_{2}$ and $a_{4}$  
(as also used in \cite{SchmYa99}),
enable one to fit all the moment constraints for $\langle \xi^{N} \rangle_\pi$.
For completeness we write explicit formulae for the optimal DA models 
\begin{eqnarray}
 \label{optG}
 \varphi_{1,2}^{\text{opt}}(x) &=& \varphi^{\rm as}(x)
 \left[1+a_2^{\text{opt1,2}} \cdot C^{3/2}_2(2x-1)
        +a_4^{\text{opt1,2}} \cdot C^{3/2}_4(2x-1) \right]\ ,\\
 a_2^{\text{opt1}}(\mu^2)&=& + 0.188\ ,\quad
 a_4^{\text{opt1}}(\mu^2)\ =\ - 0.130\ ,\quad
 \text{at}~~\lambda_q^2= 0.4~\gev{2}; \nonumber \\
 a_2^{\text{opt2}}(\mu^2)&=& + 0.126\ ,\quad
 a_4^{\text{opt2}}(\mu^2)\ =\ - 0.090\ ,\quad
 \text{at}~~\lambda_q^2= 0.5~\gev{2}, \nonumber
 \end{eqnarray} 
at $\mu^2\simeq 1~\gev{2}$.
In this way, the admissible bunches of profiles 
can be mapped into the $a_2(\mu^2),~a_4(\mu^2)$ plot 
and then can be evolved to a new normalization point
\cite{BMS01,SchmYa99}, $\mu_{S\&Y}$,
see slanted rectangles in  Fig.\ref{fig:ogurec}.

\textbf{Pion DA vs CLEO data}.
Recently, the CLEO collaboration~\cite{CLEO98} measured the 
$\gamma^{*}\gamma \to \pi^{0}$ form factor with a high accuracy.
These data sets were processed by 
Schmedding and Yakovlev (S\&Y) \cite{SchmYa99} 
using a NLO light-cone QCD SR analysis.
%%%%%%%%%%%%%%%%%%%%%%%%%%%%%%%%%%%%%%%%%%%%%%%%%%%%%%%%%%%%%%%%%%%%%%%%
%%%%% FIGURE 5: Ogurcy for lambda_q^2 = 0.4 - 0.6 GeV^2 %%%%%%%%%%%%%%%
%%%%%%%%%%%%%%%%%%%%%%%%%%%%%%%%%%%%%%%%%%%%%%%%%%%%%%%%%%%%%%%%%%%%%%%%
%%%%%%%%%%%%%%%%%%%%%%%%%%%%%%%%%%%%%%%%%%%%%%%%%%%%%%%%%%%%%%%%%%%%%%%%
They obtained the constraints to $(a_2, a_4)$ coefficients
in the form of 95\%
($2\sigma$-deviation criterium) 
and 68\% 
($1\sigma$-deviation criterium) 
confidence regions,
see ellipse-like contours in Fig.~\ref{fig:ogurec} 
with the central point $a_2^{\ast}=0.190,~a_2^{\ast}=-0.140$ 
that (approximately) corresponds to an average normalization point 
of $\mu_{S\&Y}=2.4$~GeV.

  The regions enclosed by the slanted rectangles 
  bounded by the short-dashed line and solid line in Fig.\ref{fig:ogurec}
correspond to the bunches of DA displayed in Fig.\ref{fig:DArange}(a) and (b).
The central point $a_2=0.142,~a_4=-0.087$
corresponding to the optimum profile of the $\varphi_1(x)$--bunch 
($\lambda_q^2 = 0.4~\gev{2}$ ~Fig.3(a)) at $\mu^2_{S\&Y} $ 
definitively belongs to the central S\&Y region.   
 The $\varphi_2(x)$--bunch is, however, mostly outside the central $68\%$ region
though still within the $95\%$ confidence region.
Finally, the third slanted rectangle
limited by the long-dashed contour
and shifted %along the ``diagonal'' 
to the upper left corner 
of the figure in Fig.\ref{fig:ogurec} 
corresponds to the trial bunch of NLC-DA with $\lambda_q^2=0.6~\gev{2}$. 
This value falls actually outside the standard QCD NLC-SR bounds 
in Eq.(\ref{eq:lambda}) for $\lambda_q^2$.
Remarkably, the image of this region in Fig.\ref{fig:ogurec}
lies completely outside the central region as a whole. 
Therefore, we may conclude that the CLEO data prefer the values
$\lambda_q^2=0.4,~0.5~\gev{2}$ and probably do not prefer 
the value $\lambda_q^2=0.6~\gev{2}$, 
in full agreement with previous QCD SR estimates. 
Now the conclusion 
is supported by the lattice results presented in section 3.
The quantitative details of the above qualitative discussion 
can be found in \cite{BMS01}.

\textbf{Pion DA from nondiagonal correlator. The ``Advanced" Ansatz.}
An approach to obtain directly the forms of the pion and its first resonance
DAs, by using the available smooth Ansatz for the correlation
functions  $f(\alpha)$ of the nonlocal condensates, was suggested in
papers \cite{Rad94,BM95}.
  The sum rule for these DAs
$\varphi_{\pi^{\ldots}}(x)$ %of the pseudoscalar mesons,
based on the nondiagonal correlator of the axial and
pseudoscalar currents has the vanishing perturbative density, 
\begin{eqnarray}
  \varphi_{\pi}(x)
 + \varphi_{\pi'}(x) e^{-m_{\pi'}^2/M^2}
 + \varphi_{\pi''}(x) e^{-m_{\pi''}^2/M^2}
 + \ldots
 \ =\qquad\quad
  && %\equiv  \Phi(\frac{1}{M^2},x)
 \nonumber \\
 \frac{M^2}{2}
  \left[
   \left(1 - x + \frac{\lambda_q^2}{2M^2}
   \right)f_S(x M^2)
   +
   \left(x + \frac{\lambda_q^2}{2M^2}
   \right)f_S((1-x) M^2)
  \right]   
 && \label{ar-sr}
\end{eqnarray}
only quark and mixed condensates 
appear in the theoretical part of the SR~\cite{Rad94}.
The SR results in the elegant Eq.(\ref{ar-sr})  
from the approximations both in the theoretical
(for a detailed discussion, see ~\cite{Rad94}) and
the phenomenological parts (see ~\cite{BM95}).
In virtue of the approximations, a single
correlation function  $f_S(\alpha)$ appears  in the r.h.s. of Eq.(\ref{ar-sr})
that determines the profile of $\varphi_{\pi}(x)$.

Equation (\ref{ar-sr}) demonstrates, in the
most explicit manner, an important relation: the distribution $\varphi_{\pi}(x)$
of quarks inside the pion over the longitudinal momentum
fraction $x$ (on the l.h.s.) is directly related
to the distribution $f_S(\alpha)$ over the virtuality $\alpha$ of the
vacuum quarks on the r.h.s..
Note that a similar relation was obtained in an instanton-induced
model \cite{ADT00}.

The approaches to extract $\varphi_{\pi}(x)$ with the help of SR (\ref{ar-sr}) 
were discussed in detail in \cite{Rad94,BM95}, 
here we concentrate on the final result for the profile in the case of
the Advanced ansatz (\ref{eq:A22}). 
Note only that these approaches do not provide the behaviour of the
profile in the vicinity of the end points, the reliable predictions
are expected around the mid-region.
%%%%%%%%%%%%%%%%%%%%%%%%%%%%%%%%%%%%%%%%%%%%%%%%%%%%%%%%%%%%%%%%%%%%%%%%
%%%%% FIGURE 6: Ogurcy for lambda_q^2 = 0.4 - 0.6 GeV^2 %%%%%%%%%%%%%%%
%%%%%%%%%%%%%%%%%%%%%%%%%%%%%%%%%%%%%%%%%%%%%%%%%%%%%%%%%%%%%%%%%%%%%%%%
\begin{figure}[hb]
 \myfig{\label{fig:pi-res3}}
  \noindent
   \hspace*{0.25\textwidth}
    \begin{minipage}{0.5\textwidth}\Large
     \centerline{\includegraphics[width=\textwidth]{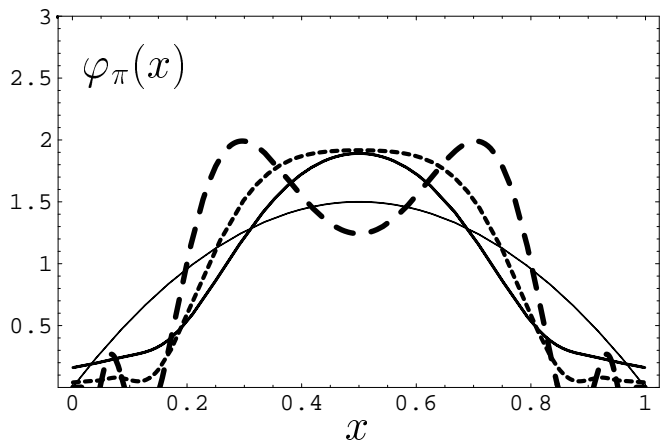}}
    \end{minipage}\vspace*{1mm}
     \caption{\footnotesize%
     Graphical representation of $\varphi_{\pi}(x)$ in the 3-resonance
     approximation. Solid line corresponds to $M^2=0.4~\gev{2}$, 
     short-dashed line -- $M^2=0.6~\gev{2}$, long-dashed line 
     -- $M^2= 1~\gev{2}$, fine line -- asymptotic DA.    
     }
\end{figure}
%%%%%%%%%%%%%%%%%%%%%%%%%%%%%%%%%%%%%%%%%%%%%%%%%%%%%%%%%%%%%%%%%%%%%%%%
\vspace*{1mm}

\noindent 
The shape of $\varphi_{\pi}(x)$ strongly depends 
on the spectrum of the resonances 
in the l.h.s. of Eq.(\ref{ar-sr})
and weaker -- on the value of $\lambda_q^2$.
For the model of equidistant, infinite number of narrow excitations 
like the ``Dirac comb" 
we have obtained the profile,
very close to the asymptotic DA~\cite{BM95}.
If we choose the spectral density in accordance with the current knowledge%
\setcounter{footnote}{1}%
\footnote{%
Only $\pi'$ and $\pi''$ are well established,
while the mass of $\pi'''$ is fixed in analogy with
other light meson spectra.},
\ie, containing only 3 radial $\pi$-excitations 
with the masses 
$m^2_{\pi'} \approx 1.3^2$~\gev{2},
$m^2_{\pi''} \approx 1.8^2$~\gev{2},
$m^2_{\pi'''} \sim 4.7$~\gev{2},
then we obtain the set of admissible profiles 
presented in Fig.\ref{fig:pi-res3}. 
It is naturally to suggest that $\varphi_{\pi}(x)$ 
is situated between these two possibilities.
This result qualitatively agrees with the profile behaviour 
for the bunches $\varphi_1(x)$ and $\varphi_2(x)$ 
in Fig.\ref{fig:DArange}.

\section{Conclusion}
We consider the admissible Ansatzes for the coordinate behaviour 
of the scalar quark nonlocal condensate
being of importance 
in QCD SRs.
These Ansatzes depend on the parameter $\lambda_q^2$ --- 
the short-distance correlation scale 
in the QCD vacuum 
that controls the corresponding coordinate behaviour.
We analyse the lattice simulation data for the scalar NLC from \cite{DDM99} 
and test two different Ansatzes.
The correlation scales are extracted following the procedure 
that includes a fit of lattice data, 
an extrapolation to the chiral limit, 
and an evolution of the obtained lattice results 
to the characteristic scale $\mu^2=1~\gev{2}$ 
of a continuum QCD.

The scale $\lambda_q^2$ thus extracted
does not visibly depend on the kind of a lattice 
(full or quenched) QCD,
as well as on the kind of the Ansatz. 
The value of the scale appears in a good agreement
with the QCD range 
$\lambda_q^2(1~\gev{2}) = (0.4 - 0.55)~\gev{2}$, 
see Figs.2--3 and Table 4 in section 3. 
The agreement looks unexpectedly good 
in view of a roughness 
of the mentioned procedure. 

The scalar condensate 
and the scale $\lambda_q^2$ 
substantially determine the shape of the pion distribution amplitude 
by means of QCD SRs. 
Both kinds of the considered Ansatzes 
lead to similar shapes of the pion DA. 
The pion DA following from QCD SR for the moments \cite{BM98,BMS01} 
is rather sensitive to the value of $\lambda_q^2$. 
The bunches of DA profiles corresponding to the $\lambda_q^2$ values 
from the QCD range do just agree with the constraints 
that follow from the CLEO experimental data~\cite{BMS01}.

This and previous~\cite{MR92,Rad94,BM95,BM98,BMS01}
considerations demonstrate a close link 
between the correlation scale 
in the QCD vacuum and the shape of the pion DA. 
%By this reason the results of lattice condensate measurements 
%agree at the very end 
%with the CLEO experimental data.
\vspace{1cm}

\textbf{Acknowledgments}\\
This work was supported in
part by the Russian Foundation for Fundamental Research
(contract 00-02-16696), INTAS-CALL 2000 N 587, 
the Heisenberg--Landau Program (grants 2000-15 and 2001-11),
and the INFN--JINR Cooperation Program.
We are grateful to A.~Di~Giacomo, A.\ Dorokhov, 
M.~D'Elia, E.~Meggiolaro and O.~Teryaev for discussions.
We are indebted to Prof.~A.~Di~Giacomo
for the warm hospitality at Pisa University where this
work was completed.

\begin{appendix}
\vspace*{12mm}
\appendix
\hspace*{2mm}{\Large \bf Appendix}
\vspace{-3mm}
%%%%%%%%%%%%%%%%%%%%%%%%%%%%%%%%%%%%%%%%%%%%%%%%%%%%%%%%%%%%%%%%%%%%%
%%%%%%%%%%%%%%%%%%%%%%%%%%%%%%%%%%%%%%%%%%%%%%%%%%%%%%%%%%%%%%%%%%%%%
\section{Basic set of local condensates}
 \renewcommand{\theequation}{\thesection.\arabic{equation}}
  \label{app-A}\setcounter{equation}{0}
%%%%%%%%%%%%%%%%%%%%%%%%%%%%%%%%%%%%%%%%%%%%%%%%%%%%%%%%%%%%%%%%%%%%%
%%%%%%%%%%%%%%%%%%%%%%%%%%%%%%%%%%%%%%%%%%%%%%%%%%%%%%%%%%%%%%%%%%%%%
Here we write down the explicit expressions~\cite{Gro94}
for the condensates $Q^i$
that determine the first and second Taylor expansion coefficients 
in (\ref{eq:QS}), 

\begin{eqnarray}
Q^7_{(0)} &=& 3Q^7_1-\frac32Q^7_2-3Q^7_3+Q^7_4,  \\
Q^6_{(m)} &=& 2Q^6+3mQ^5-3m^3Q^3,  \\
Q^8_{(0)}&=& 5A-{\textstyle\frac52}Q^8_2+{\textstyle\frac14}Q^8_3
-{\textstyle\frac12}Q^8_4-3Q^8_5+5Q^8_6, \\
Q^7_{(m)} &=& 5(3Q^7_1-Q^7_2-3Q^7_3+Q^7_4)
 -15m(Q^6+mQ^5-m^3Q^3),
\end{eqnarray}
where the quark condensate basis was chosen in the form 
\begin{equation} 
  Q^3 = \qq{}\,,\quad
  Q^5 = ig\qq{\left(G_{\mu\nu}\si_{\mu\nu}\right)},
    \quad
  Q^6 = \va{t^aJ^a_\mu t^bJ^b_\mu},
\end{equation}
for the condensates of lowest dimensions, and 
\begin{eqnarray}  \label{eq:Q^7}                                                               
Q^7_1 & = & <\bar q G_{\mu\nu}G_{\mu\nu} q>,\ \ \ \                 
Q^7_2 = i <\bar q G_{\mu\nu}\tilde G_{\mu\nu} \gamma_5 q>, \\                                                               
Q^7_3 & = & <\bar q G_{\mu\lambda}G_{\lambda\nu} \sigma_{\mu\nu} q>,
\ \ \ \                                                             
Q^7_4 = i <\bar q D_{\mu}J_\nu \sigma_{\mu\nu} q>,                  
\nn\end{eqnarray}  
for the condensates of dimensions 7. 
The basis elements of  dimension 8, $Q^8_i$ and $A $, 
enter into the expansion only of the vector NLC (\ref{eq:QV})
that is not analyzed here. 
For this reason, we do not show it here 
and refer the reader to article \cite{Gro94}, Eqs.(3.10-3.11).

%%%%%%%%%%%%%%%%%%%%%%%%%%%%%%%%%%%%%%%%%%%%%%%%%%%%%%%%%%%%%%%%%%%%%  
%%%%%%%%%%%%%%%%%%%%%%%%%%%%%%%%%%%%%%%%%%%%%%%%%%%%%%%%%%%%%%%%%%%%  
  \section{Details of Pisa lattice simulations}                          
   \renewcommand{\theequation}{\thesection.\arabic{equation}}            
    \label{app-B}\setcounter{equation}{0}                                   
%%%%%%%%%%%%%%%%%%%%%%%%%%%%%%%%%%%%%%%%%%%%%%%%%%%%%%%%%%%%%%%%%%%%                                                              
%%%%%%%%%%%%%%%%%%%%%%%%%%%%%%%%%%%%%%%%%%%%%%%%%%%%%%%%%%%%%%%%%%%%%  
  For the full QCD case (with dynamical fermions),                        
the nonlocal condensates were measured  on a $16^3 \times 24$ lattice                                                                          
 at $\beta = 5.35$                                                                        
($\beta = 6/g^2$, where $g$ is the coupling constant) 
and two different values of the dynamic  quark mass: 
$a \cdot m_q = 0.01$ and $a \cdot m_q = 0.02$
with the following values for the lattice spacing:
\begin{eqnarray}
 a(\beta = 5.35) &\simeq& 0.101~\text{fm},~~\text{for}~ a \cdot m_q = 0.01 ~;
\nonumber \\
 a(\beta = 5.35) &\simeq& 0.120~\text{fm},~~\text{for}~ a \cdot m_q = 0.02 ~.
\label{ref:f}
\end{eqnarray}
For the quenched QCD case the measurement was performed on a $16^4$
lattice at $\beta = 6.00$, with the quark mass $a \cdot m_q = 0.01$ for
constructing the external--field quark propagator, and also at $\beta = 5.91$,
with the quark mass $a \cdot m_q = 0.02$. In both the cases, the value of $\beta$
was chosen in order to have the same physical scale as in full QCD
at the corresponding quark masses, thus allowing a direct comparison
between the quenched and the full theory.
In the quenched case, the lattice spacing is approximately~\cite{DDM99}:
\begin{eqnarray}
 a^{(YM)} (\beta = 6.00) &\simeq& 0.103~\text{fm},~~\text{for}~ a \cdot m_q = 0.01, 0.05, 0.1~;
\nonumber \\
 a^{(YM)} (\beta = 5.91) &\simeq& 0.120~\text{fm},~~\text{for}~ a \cdot m_q = 0.02~.
\label{ref:q}
\end{eqnarray}
\end{appendix}

\markboth{References}{References}
\newcommand{\noopsort}[1]{} \newcommand{\printfirst}[2]{#1}
 \newcommand{\singleletter}[1]{#1} \newcommand{\switchargs}[2]{#2#1}


\begin{thebibliography}{40}

\bibitem{DDM99}
 M. D'Elia, A. {\uppercase{d}i}~Giacomo, and E. Meggiolaro,  
  Phys. Rev. D59  (1999)  054503.
  %%CITATION = PHRVA,D59,054503;%%.
%1
\bibitem{MR86}
 S.~V. Mikhailov and A.~V. Radyushkin, 
  JETP Lett. 43  (1986)  712;
   %%CITATION = JTPLA,43,712;%%.
\\
  Sov. J. Nucl. Phys. 49  (1989)  494. 
   %%CITATION = SJNCA,49,494;%%.
%2
\bibitem{BR91}
 A.~P. Bakulev and A.~V. Radyushkin, 
  Phys. Lett. B271  (1991)  223.
   %%CITATION = PHLTA,B271,223;%%.
%3
\bibitem{MR92}
 S.~V. Mikhailov and A.~V. Radyushkin,
  Phys. Rev. D45  (1992)  1754.
   %%CITATION = PHRVA,D45,1754;%%.
%4
\bibitem{Rad94}
 A.~V. Radyushkin,
 arXiv:hep-ph/9406237.
  %%CITATION = HEP-PH 9406237;%%.
%5
\bibitem{BM94}
 A.~P. Bakulev and S.~V. Mikhailov,  
  JETP Lett. 60  (1994)  150;
   %%CITATION = HEP-PH 9406216;%%.
\\
in *Vladimir 1994, Quarks '94* 574-580.
%6
\bibitem{BM95}
 A.~P. Bakulev and S.~V. Mikhailov, 
  Z. Phys. C68  (1995)  451; 
   %%CITATION = HEP-PH 9412366;%%.
\\
  Mod. Phys. Lett. A11  (1996)  1611. 
   %%CITATION = HEP-PH 9512432;%%.
%7
\bibitem{BM98}
 A.~P. Bakulev and S.~V. Mikhailov,
  Phys. Lett. B436  (1998)  351.
   %%CITATION = PHLTA,B436,351;%%.
%8
\bibitem{CLEO98}
 J. Gronberg \textit{ et~al.},
  Phys. Rev. D57  (1998)  33.
   %%CITATION = PHRVA,D57,33;%%.
%9
\bibitem{BMS01}
 A.~P. Bakulev, S.~V. Mikhailov, and N.~G. Stefanis,
  Phys. Lett. B508  (2001)  279.
   %%CITATION = HEP-PH 0103119;%%.
%10
\bibitem{Gro94}
 A.~G. Grozin, 
  Int. J. Mod. Phys. A10  (1995)  3497.
   %%CITATION = HEP-PH 9412238;%%.
%11
\bibitem{SVZ}
 M.~A. Shifman, A.~I. Vainshtein, and V.~I. Zakharov,
  Nucl. Phys. B147  (1979)  385; %%CITATION = NUPHA,B147,385;%%.
448;  %%CITATION = NUPHA,B147,448;%%.
519.  %%CITATION = NUPHA,B147,519;%%.
%12
\bibitem{BI82}
 V.~M. Belyaev and B.~L. Ioffe, 
  Sov. Phys. JETP 57  (1983)  716;
   %%CITATION = SPHJA,57,716;%%.
\\
 A.~A. Ovchinnikov and A.~A. Pivovarov,
  Sov. J. Nucl. Phys. 48  (1988)  721.
   %%CITATION = SJNCA,48,721;%%.
%13
\bibitem{piv91}
 A.~A. Pivovarov,
  Bull. Lebedev Phys. Inst. 5  (1991)  1.
   %%CITATION = SPLRD,5,1;%%.
%14
\bibitem{DEM97}
 A.~E. Dorokhov, S.~V. Esaibegian, and S.~V. Mikhailov,
  Phys. Rev. D56  (1997)  4062.
   %%CITATION = PHRVA,D56,4062;%%.
\\
%\bibitem{PW96}
 M.~V. Polyakov and C. Weiss,
  Phys. Lett. B387  (1996)  841.
   %%CITATION = PHLTA,B387,841;%%.
%15
\bibitem{Rad91}
 A.~V. Radyushkin, 
  Phys. Lett. B271  (1991)  218.
   %%CITATION = PHLTA,B271,218;%%.
%16
\bibitem{MR90}
 S.~V. Mikhailov and A.~V. Radyushkin, 
  Sov. J. Nucl. Phys. 52  (1990)  697.
   %%CITATION = SJNCA,52,697;%%.
%17
\bibitem{GroY92}
 A.~G. Grozin and Oleg~I. Yakovlev, 
  Phys. Lett. B291  (1992)  441.
     %%CITATION = PHLTA,B291,441;%%.
%18
\bibitem{Piv96}
A.~A. Pivovarov,
 Phys. Atom. Nucl. 59 (1996) 891.
  %%CITATION = YAFIA,59,930;%%
%19   
\bibitem{DEJM00}
 H.~G. Dosch, M. Eidem{\"u}ller, M. Jamin, and E. Meggiolaro, 
  JHEP 07  (2000)  023.
   %%CITATION = JHEPA,0007,023;%%.
%20
\bibitem{Meg99}
 E. Meggiolaro, 
  Nucl. Phys. Proc. Suppl. 83  (2000)  512.
   %%CITATION = NUPHZ,83,512;%%.
%21
\bibitem{Alt93}
 R. Altmeyer \textit{ et~al.}, 
  Nucl. Phys. B389  (1993)  445.
   %%CITATION = NUPHA,B389,445;%%.
%22
\bibitem{Mor84}
 A.~Yu. Morozov, 
  Sov. J. Nucl. Phys. 40  (1984)  505.
   %%CITATION = SJNCA,40,505;%%.
%23
\bibitem{KS87}
 M.~Kremer and G.~Schierholz,
  Phys. Lett. B194  (1987) 283.
   %%CITATION = PHLTA,B194,283;%%
%24 
\bibitem{CZ84}
 V.~L. Chernyak and A.~R. Zhitnitsky, 
  Phys. Rept. 112  (1984)  173.
   %%CITATION = PRPLC,112,173;%%.
%25
\bibitem{CZ77}
 V.~L. Chernyak and A.~R. Zhitnitsky, 
  JETP Lett. 25  (1977)  510.
   %%CITATION = JTPLA,25,510;%%.
%26
\bibitem{ERBL79}
 G. P. Lepage and S. J. Brodsky,
 Phys. Lett. B 87 (1979) 359;
 %%CITATION = PHLTA,B87,359;%%.
 Phys. Rev. D 22 (1980) 2157;
 %%CITATION = PHRVA,D22,2157;%%.
\\
 A. V. Efremov and A. V. Radyushkin,
 Theor. Math. Phys. 42 (1980) 97;
 %%CITATION = TMPHA,42,97;%%.
 Phys. Lett. B 94 (1980) 245.
 %%CITATION = PHLTA,B94,245;%%.
%27
\bibitem{SchmYa99}
 A. Schmedding and O. Yakovlev, 
  Phys. Rev. D62  (2000)  116002.
   %%CITATION = PHRVA,D62,116002;%%.
%28
\bibitem{ADT00}
I. V. Anikin, A. E. Dorokhov, L. Tomio,
 Phys. Lett. B475  (2000)  361.
   %%CITATION = PHLTA,B475,361;%%.
%29   
\end{thebibliography}
\end{document}